# Polyurea-Graphene Nanocomposites – the Influence of Hard-Segment Content and Nanoparticle Loading on Mechanical Properties

**Demetrios A. Tzelepis [1,2], Arman Khoshnevis [3], Mohsen Zayernouri [4,5] and Valeriy V. Ginzburg [6*]**

1  Department of Chemical Engineering and Materials Science, Michigan State University, East Lansing, Michigan, USA; email tzelepi1@msu.edu
2  Materials Division, US-Army, Ground Vehicle System Center, Warren, Michigan, USA
3  Department of Mechanical Engineering, Michigan State University, East Lansing, Michigan, USA; email khoshne1@msu.edu
4  Department of Mechanical Engineering, Michigan State University, East Lansing, Michigan, USA; email zayern@msu.edu
5  Department of Statistics and Probability, Michigan State University, East Lansing, Michigan, USA
6  Department of Chemical Engineering and Materials Science, Michigan State University, East Lansing, Michigan, USA; email ginzbur7@msu.edu
*  Correspondence: ginzbur7@msu.edu

Abstract: Polyurethane and polyurea-based adhesives are widely used in various applications, from automotive to electronics to medical. The adhesive performance depends strongly on its composition, and developing the formulation—structure—property relationship is crucial to making better products. Here, we investigate the dependence of the linear viscoelastic properties of polyurea nanocomposites, with IPDI-based polyurea (PUa) matrix and exfoliated graphene nanoplatelet (xGnP) fillers, on the hard segment weight fraction (HSWF) and the xGnP loading. We characterize the material using scanning electron microscopy (SEM) and dynamical mechanical analysis (DMA). It is found that changing HSWF leads to a significant variation in the stiffness of the material, from about 10 MPa for the 20% HSWF to about 100 MPa for the 30% HSWF to about 250 MPa for the 40% HSWF polymer (as measured by the tensile storage modulus at room temperature). The effect of the xGNP loading is significantly more limited and is generally within experimental error, except for the 20% HSWF material where the xGNP addition leads to about 80% increase in stiffness. To correctly interpret the DMA results, we developed a new physics-based rheological model for the description of the storage and loss moduli. The model is based on the fractional calculus approach and successfully describes the material rheology in a broad range of temperatures (-70°C -- +70°C) and frequencies (0.1 – 100 s$^{-1}$), using only six physically meaningful fitting parameters for each material. The results provide guidance for the development of nanocomposite PUa-based materials.

## 1. Introduction

Polyurethanes (PU), polyureas (PUa), and poly(urethaneureas) (PUU), represent a class of polymers with a wide variety of applications.[1–4]   Understanding the structure-property-performance relationship is critical in designing materials for specific applications. Critical parameters include chemical structure of polymer constituents, extent of hydrogen bonding, and volume fraction of hard and soft segments.[5–12]   In general, these classes of polymers are a reaction between polyisocyanate (typically a diisocyanate) and a polyol in the case of pure PU and a polyamine in case of pure PUa and both polyol and polyamine for PUU.   In PUa, which will be the focus of this paper, the reaction of the diisocyanate and a polyamine form the hard segments which have strong bidentate hydrogen bonds. [1]   The PUa can then be thought of as a multiblock copolymer in which the soft segment blocks alternate with the hard segment blocks. The strong hydrogen bonding within the hard segments drives microphase separation from the soft segments, resulting in a two-phase system -- a percolated hard phase, consisting entirely of the hard segments, and a soft phase consisting of the soft segments along with small amounts of non-percolated hard segments. [10] This microphase separation is similar to that of classical block copolymers where various soft-crystalline phase (spherical, cylindrical, lamellar, etc.) are seen for different values of the composition, $f$, and the segregation strength, $\chi N$. [13–15] The morphology of the polymer, especially the total volume and the connectivity of the hard phase, has a decisive impact on the overall material properties (mechanical, transport, and thermophysical).[10,16–18]

Much of the PU and PUa literature concentrated on the linear elasticity, and especially the "temperature sweep" dynamical mechanical analysis (DMA), where the storage and loss moduli are measured as a function of temperature at constant frequency (usually 1Hz). However, many applications (automotive, electronics, etc.) require a good understanding of material performance in a wide range of temperatures and frequencies/strain rates. Thus, recently Tzelepis et al.[19] used both temperature-sweep and frequency-sweep DMA to study the properties of PUa elastomers with different hard segment weight fractions (HSWF). It was shown that the studied PUa materials obeyed the time-temperature superposition (TTS) principle. (We note that the application of TTS to PU and PUa was discussed earlier, e.g., by Velankar and Cooper [20] and Ionita et al. [21], but whether it is universally applicable to all PU and PUa is still uncertain). The TTS shift factor, $a_T$, was successfully described using the TS2 function[22] that combines Arrhenius temperature dependence at high temperatures with a strong but non-divergent increase near the glass transition temperature. The storage and loss modulus master curves showed broad transition regions, indicating a wide distribution of relaxation times. Tzelepis et al. found that such a distribution is well-described by the so-called fractional Maxwell model (FMM) [23–27] – or, to be more precise, a sum of two fractional Maxwell gels (FMG), with one FMG element describing the continuous soft phase (with dispersed hard domains and dissolved hard segments) and the second FMG element representing the percolated hard phase. The plateau modulus of the first element was found to be nearly independent of the HSWF, while the plateau modulus of the second element was a strong function of HSWF, consistent with earlier experiments and theories.[10]

In this study we extend our previous work to investigate a set of PUa nanocomposites with exfoliated graphene nanoplatelets (xGnP), varying both HSWF and the xGnP weight fraction. The use of nanofillers such as clay, talc, graphene, and graphene oxide to modify the properties of polymers have been widespread since at least 1980s. [28–39] The fillers are expected to significantly increase modulus and strength of the material relative to the "neat" matrix polymer. For high-aspect ratio nanoplatelets in rubbery polymers, the "reinforcement factor" (RF) – defined as the ratio of the nanocomposite modulus to the matrix modulus – can be as high as 2-4 at particle loadings of 1-4 weight percent.[40,41] Multiple models have been developed to predict the reinforcement in simple two-component nanocomposites. [40–42] In general, the stiffness of the material increases strongly at the beginning, but often stays constant or even decreases as the filler loading is increased further – this is typically ascribed to the onset of nanofiller aggregation. Obviously, the problem becomes even more challenging when the matrix itself is multicomponent, like segmented polyurea. Are the nanofillers simply interacting with the pre-set domain nanostructure? Or, are they modifying the arrangement of the hard domain itself – perhaps by nucleating their formation or by linking multiple domains? Here, we will attempt to address this problem by preparing multiple PUa-xGnP nanocomposites and investigating their structure and linear viscoelasticity. Starting with the three neat PUa materials (20, 30, and 40 percent hard segment), we add the xGnP nanofillers, with the xGnP weight percentage (wt %) varied from 0 to 1.5 wt % with increments of 0.5 wt %. We expected that this experimental design would capture the main reinforcement effect due to the nanofillers. On the one hand, reinforcement effects are unlikely to be significant at loadings below 0.5 wt %, based on many earlier polymer nanocomposite studies (see, e.g., Pinnavaia and Beall [43]). On the other hand, as it will be seen later, at loadings 1.5 wt % and above, the reinforcement effects diminish, possibly due to the nanoparticle interactions and transition from fully to partially exfoliated morphology. For all twelve materials (neat PUa and nanocomposites), we measured the linear viscoelasticity and successfully fit it with the two-FMG model. The model parameters were then used to elucidate the structural details of the material and provide guidance for the impact of the design parameters (HSWF and wt % xGnP) on the nanocomposite properties.

## 2. Materials and Methods

### 2.1. Polymer Synthesis

The synthesis of the PUa-Neat materials is described in detail in our earlier paper. [19] We used isophorone diisocyanate (IPDI)-Vestanat from Evonik Corporation, Jeffamine T5000 and D2000 polyetheramines from the Huntsman Corporation, and the diethyltoluene diamine (DETDA) (Lonzacure) chain extender from Lonza. Toluene was purchased from Fisher Scientific. All the materials were employed in our research "as received" with no further processing. The formulations for the three neat PUa-s having hard segment weight fractions of 20%, 30%, and 40%, are given in Table 1. We made polyurea prepolymer (A-side) by placing IPDI in the reactor, then adding toluene to prevent any possible gelling. Next, the amine blend for the prepolymer (comprised of Jeffamine D2000 and T5000) was mixed for 5 minutes in a separate 250 ml beaker at room temperature, subsequently degassed for approximately 10 minutes, and added to the IPDI-toluene mixture. Similarly, the amine blend for the B-side, comprised of Jeffamine D2000 and Lonzacure DETDA, was mixed for 5 minutes in separate 250 ml beaker followed by vacuum degassing for approximately 10 minutes and then poured into a separate additional funnel. The reactor was assembled and then a vacuum was drawn for five minutes, followed by the addition of $N_2$ gas at 0.3-0.4 L/min flow rate. The reactor temperature was increased to 80 ℃ and then A-side amine blend was added dropwise under mechanical stirring at 120 RPMs. The mixture was subsequently stirred for another hour at 80 ℃. Afterwards, the reactor was cooled to 0 ℃ and the B-side amine blend was added dropwise, with mechanical stirring maintained at 120 RPMs. Once all the B-side was added, the contents of the reactor were transferred into a 600 ml beaker, degassed for 5 minutes, and poured into molds. The molds were kept at room temperature for 24 hours to allow for gelation and solvent evaporation. After 24 hours the samples were placed in an oven at 40 ℃, for 12-24 hours, to accelerate the solvent evaporation. Curing of the PUa was then completed at 60℃ for 72 hours.

For each of the three PUa formulations described in Table 1, four nanocomposites were then prepared, with the xGnP weight percentage (wt %) varied from 0 to 1.5 wt % with increments of 0.5 wt %. For all the PUa-GnP nanocomposites, the process was identical to that for the neat systems with the following additional steps. Exfoliated nano-graphene (grade R-10, obtained from XG Sciences) was heat-treated at 400 ℃ for 1 hour and allowed to furnace-cool. The required amount of xGnP's were placed in a 500 ml beaker, 190 ml of toluene was added to the beaker and the slurry was simultaneously mechanically stirred and sonicated. The mechanical stirring was accomplished by magnetic stirring at 200 rpm. The sonication was accomplished using a Qsonica sonicator. The amplitude was set to 20 and the process time was set to 30 minutes with a pulse time of 10 seconds on and 10 seconds off. The temperature of the slurry never exceeded 32 ℃, and the total run time was ~1 hour. The total amount of energy input was 38,610 J. The weight of xGnP added to the formulation is summarized as follows: for the 0.5 wt% xGnP formulations -- 1.02 g of xGnP; for the 1.0 wt % xGnP formulations -- 2.04 g of xGnP; and for the 1.5 wt% xGnP formulation -- 3.06 g of xGnP.

Table 1. Summary of the constituents used in the synthesis of the model PUa - Neat. (Adapted with permission from Ref. [19]

|  | Component | IPDI-2k-20HS | IPDI-2k-30HS | IPDI-2k-40HS |
|---|---|---|---|---|
| Isocyanate Prepolymer (A-Side) | IPDI | 30.8 g | 41.6 g | 52.1 g |
| | T5000 | 14.5 g | 12.1 g | 10.1 g |
| | D2000 | 57.1 g | 48.5 g | 40.4 g |
| | Toluene | 82.7 g (95 ml) | 165.3 g (190 ml) | 208.8 g (240 ml) |
| | %NCO | 8.7 % | 12.9 % | 15.3 % |
| Amine Blend (B-Side) | DETDA | 10 g | 19.4 g | 29.3 g |
| | D2000 | 90 g | 80.6 g | 70.7 g |

## 2.2 PUa-xGnP Characterization

The surface chemistry of the top 50-80 Å is determined with X-Ray Photoelectron Spectroscopy (XPS). The measurements were performed using a PHI 5400 ESCA system. The base pressure of the instrument was less than $10^{-8}$ Torr. A 1 cm 2 sample was mounted onto the sample holder with double sided copper tape. The X-Ray was a monochromatic Al source with a take-off angle of 45 degrees. Two types of scans were performed for each sample: a survey scan from 0-1100 eV taken with a pass energy of 187.85 eV and regional scans of each element at a pass energy of 23.70 eV. Data was fit using the CASA XPS software package.

xGnP particle size and dispersion was characterized using a Hitachi 3700 SEM. The acceleration was set to 5 keV to minimize charging effects. A 2-to-3 nanometer-thick gold coating was sputtered, using a Quorum Q150R sputter coater. Geometric measurements of the xGnP were made utilizing PCI software

Dynamic mechanical analysis was conducted using a TA instruments RSA-G2 instrument. The curing of the polymer was determined by measuring the change in storage modulus with respect to time. All film samples were loaded in tension. Temperature sweeps, at a rate of 3 °C/min, were conducted from -95 °C to a max temperature depending on the polyurea formulation hard segment content. Six repeats per formulation were run for the temperature sweeps. The reference temperature for each material was set to equal its glass transition temperature, defined as the maximum of the loss modulus (see Table S1). All TTS shifts were completed with TA instruments TRIOS software package.

## 2.3. Modeling

### 2.3.1. Fractional-order Maxwell Gel Model

The Fractional Maxwell Model (FMM) can be employed in developing constitutive models for both soft solids and complex fluids. The FMM consists of two spring-pot elements in series, which describe the complex modulus, $E^*$, as presented in Equation (1). [25]

$$\frac{E^*(\omega)}{E_0} = \frac{(i\omega\tau_c)^\alpha}{1+(i\omega\tau_c)^{\alpha-\beta}} \quad (1)$$

Here, $E_0$ represents the characteristic modulus, $\tau_c$ denotes the characteristic relaxation time, and both $\alpha$ and $\beta$ are fractional-order power-law exponents. The storage, $E'$, and loss, $E''$, moduli are obtained by splitting the complex modulus into its real and imaginary components, respectively defined as

$$\frac{E'(\omega)}{E_o} = \frac{(\omega\tau_c)^\alpha \cos\left(\frac{\pi\alpha}{2}\right) + (\omega\tau_c)^{2\alpha-\beta} \cos\left(\frac{\pi\beta}{2}\right)}{1 + (\omega\tau_c)^{\alpha-\beta} \cos\left(\frac{\pi(\alpha-\beta)}{2}\right) + (\omega\tau_c)^{2(\alpha-\beta)}}$$ (2a)

$$\frac{E''(\omega)}{E_o} = \frac{(\omega\tau_c)^\alpha \sin\left(\frac{\pi\alpha}{2}\right) + (\omega\tau_c)^{2\alpha-\beta} \sin\left(\frac{\pi\beta}{2}\right)}{1 + (\omega\tau_c)^{\alpha-\beta} \cos\left(\frac{\pi(\alpha-\beta)}{2}\right) + (\omega\tau_c)^{2(\alpha-\beta)}}$$ (2b)

Within the FMM framework, one possible special case that can occur is the Fractional Maxwell Gel (FMG) denoted by $\beta$ being set to 0, which models the material's elastic behavior past the gel point.

In our model of interest, two FMG elements are arranged in parallel, representing the *soft-phase* matrix and percolated *hard phase* of polyurea, as illustrated in Figure 1. Given the relatively low mass fraction of the added nanoparticles (only up to 1.5%), we assume that they end up dispersed within the two phases and not forming a new phase by themselves. Thus, no additional parallel branch is introduced, consistent with our prior work.[19] Consequently, each polymer is characterized by six model parameters, encompassing two characteristic moduli (often called plateau modulus) ($E_{0,1}$ and $E_{0,2}$), two relaxation characteristic times ($\tau_{c,1}$ and $\tau_{c,2}$), and two power-law exponents ($\alpha_1$ and $\alpha_2$). We expect these parameters – especially those related to the percolated hard phase (FMG2) – to depend on the material formulation, including HSWF and xGnP loading.

To integrate the DMA data across various temperatures, the time-temperature superposition (TTS) principle is employed and the shift factor, denoted as $a_T$, is assumed to hold the same for both the soft and hard phases. As a result, the master curves can be described through the subsequent equations,

$$E'(x) = \sum_{k=1}^{2} E_{0,k} \frac{(x\tau_k)^\alpha \cos\left(\frac{\pi\alpha}{2}\right) + (x\tau_k)^{2\alpha}}{1 + (x\tau_k)^\alpha \cos\left(\frac{\pi\alpha}{2}\right) + (x\tau_k)^{2\alpha}}$$ (3a)

$$E''(x) = \sum_{k=1}^{2} E_{0,k} \frac{(x\tau_k)^\alpha \sin\left(\frac{\pi\alpha}{2}\right)}{1 + (x\tau_k)^\alpha \cos\left(\frac{\pi\alpha}{2}\right) + (x\tau_k)^{2\alpha}}$$ (3b)

where $x = a_T\omega$. The equation for the shift factor as a function of temperature is discussed next.

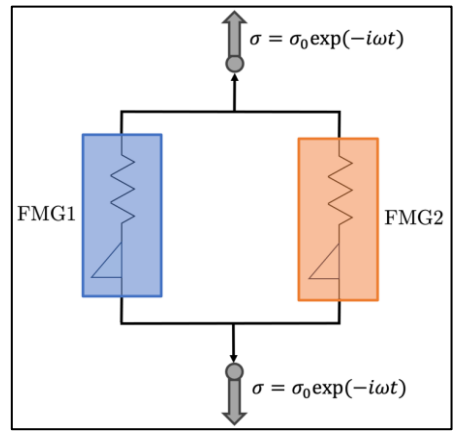

Figure 1. Schematic illustration of the two FMGs employed to model the polyurea. FMG1 represents the filled soft phase, whereas FMG2 corresponds to the percolated hard phase. No additional FMG element is considered for modeling nano-particles. Each FMG comprises an elastic spring and a spring-pot in a series arrangement.

### 2.3.2. The Shift Factor and the Two State, Two (time) Scale (TS2) Model

In previous paper, [19] we applied three different functional forms to describe the shift factor -- the Arrhenius, the Williams-Landel-Ferry (WLF),[44] and the TS2 [22]functions. The TS2 model describes the glass transition as the transition between the high-temperature and low-temperature Arrhenius regions,

$$ln(a_T) \equiv ln\left(\frac{\tau[T]}{\tau[T_0]}\right) = \frac{E_1}{RT} +$$
$$\frac{E_2 - E_1}{RT}\left(\frac{1}{1 + exp\left\{\frac{\Delta S}{R}\left(1 - \frac{T*}{T}\right)\right\}}\right) - \frac{E_1}{RT_0} +$$
$$\frac{E_2 - E_1}{RT_0}\left(\frac{1}{1 + exp\left\{\frac{\Delta S}{R}\left(1 - \frac{T*}{T_0}\right)\right\}}\right) \tag{4}$$

where $E_1$ and $E_2$ are activation energies (in J/mol), $\Delta S/R$ is the dimensionless transition entropy between the solid and liquid states of matter, $T*$ is the transition temperature (K) (typically, $T* \approx T_g$), and $T_o$ is the reference temperature of the TTS shifts. Equation (4) was shown to successfully describe the TTS of neat PUa polymers in the temperature range between -70 °C and +70 °C [19]and thus will be utilized here as well.

### 2.2.5. Optimization of the FMG Parameters

A global particle-swarm optimization (PSO) algorithm [45] is utilized to infer the fitting parameters in the two-FMG branches, depicted in Figure 1. This is the same method that was employed in our previous paper. [19] Each optimization run maintains a constant population size of $N_{pop} = 200$ and performs $N_{it} = 6000$ iterations. Given the stochastic nature of the PSO algorithm, 50 optimization runs are conducted, and the expected values and standard deviation for each parameter of materials are reported.

The following parameter ranges are considered for all samples (20%, 30%, and 40% HS, with 0, 0.5%, 1%, and 1.5% GnP). The characteristic moduli are confined to the range $0 \le E_{0,1} \le 10^4$ MPa, and $0 \le E_{0,2} \le 10^3$ MPa, the characteristic times are confined to the range $10^{-3}\,s \le \tau_{c,1(2)} \le 10^2\,s$, and the fractional power-law exponents $\alpha_1$ and $\alpha_2$ span from 0 to 1.

Equation (5) establishes a scalar multi-objective cost function via a weighted summation for the simultaneous fitting of storage and loss moduli.

$$\min_\theta \omega_1 f_1(\theta) + \omega_2 f_2(\theta) \tag{5}$$

In this equation, $\theta$ denotes the vector of fitting parameters, with $\omega_1 = 1/2$, $\omega_2 = 1/2$, and the cost functions corresponding to both moduli are provided as follows,

$$f_1(\theta) = \sum_{i=1}^{N_d}\left(\text{Log}\left(\frac{E'_{exp}}{E'_{model}}\right)\right)^2 ,$$
$$f_2(\theta) = \sum_{i=1}^{N_d}\left(\text{Log}\left(\frac{E''_{exp}}{E''_{model}}\right)\right)^2 \tag{6}$$

In these expressions, $N_d$ is the number of data points where our model is evaluated. The decision to employ a logarithmic difference between experimental data and model predictions arises from the significant variations in orders of magnitude for the storage and loss moduli across decades of frequency ranges. Moreover, the quality of the two-branch FMG model fits if assessed by the relative error as defined in Equation (7).

$$error =$$
$$\frac{\omega_1 f_1(\theta) + \omega_2 f_2(\theta)}{\omega_1 \sum_{i=1}^{N_d}\left(\text{Log}(E'_{exp})\right)^2 + \omega_2 \sum_{i=1}^{N_d}\left(\text{Log}(E''_{exp})\right)^2}. \tag{7}$$

Both the two-branch FMG model and PSO codes – similar to our previous paper [19] -- were developed in MATLAB R2021b and executed in ICER MSU HPCC system with 1 node, 24 CPU, and 48 GB RAM.

Once again, we selected the GRG non-linear engine and imposed the following constraints: 1) E1 < 130 kJ/mol, 2) E2 < 350 kJ/mol, 3) $\Delta$S/R < 25, and 4) T* < 350 K. The reference temperature (T$_o$) was set to match the glass transition temperature, defined as the maximum of the loss modulus (-60 ± 5 °C, for more details see Table S1). The minimization function utilized is the average absolute value of the difference in the natural logarithm of the experimental and model shift factors, as defined in Equation (8).

$$\overline{error} = \left| ln\left(aT_{exp}\right) - ln\left(aT_{model}\right) \right| \qquad (8)$$

This concludes the discussion of materials and methods; we now turn to the results.

## 3. Results

### 3.1 Experimental Results

Scanning electron microscopy (SEM) was used to determine the effect of sonication on the xGnP. Figure S1 shows SEM micrographs at various magnifications. Estimates of the particle diameter were made by measuring the longest axis of the platelets as shown in Figure S2. The average particle diameter before sonication was 15.4 +/- 6.3 µm (1σ). The average particle diameter after sonication was 15.0 +/- 4.5 µm (1σ). No change was seen in the morphology of the xGnP. The xGnP remained exfoliated throughout the sonication process and retained their shape and aspect ratio. Given that the technical data sheet for the R10 grade specifies an average particle diameter size of approximately 10 µm, and accounting for the fact that the platelets in the images are at various angles, we conclude that there is no difference between the as received and after sonicated xGnP.

X-ray photoelectron spectroscopy (XPS) was used to evaluate the surface chemistry of the xGnP after heat treatment and after sonication. Figure 2 shows a survey of both a heat-treated sample and heat treated and sonicated sample. Both spectra showed two peaks. The first at 281.6 eV, and 282.4 eV. heat-treated & sonicated respectively, is associated with the C 1s position. The second at 530.4 eV, for both heat-treated & sonicated, is associated with the O 1s position. The atomic concentration was estimated and is tabulated in Figure 2.

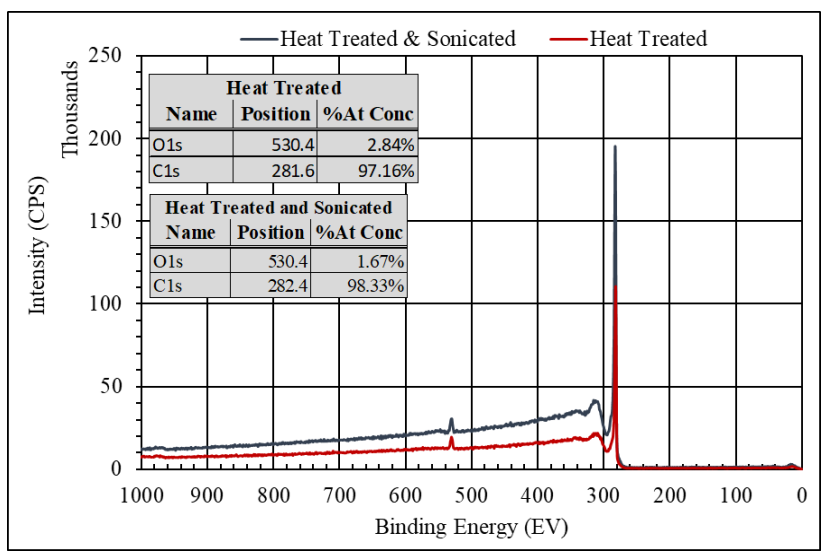

Figure 2. XPS survey of the xGnP after heat treatment (red spectrum), and after heat treatment and sonication. The sonication did not cause any change to the surface chemistry of the xGnP.

The atomic percent of the C is significantly higher than O for both heat-treated and heat-treated and sonicated samples. This is consistent with the expectation that the majority of the xGnP is carbon with very little oxygen-based functionalization on the edges of the basal plane. The approximate 1% difference seen between the two treatments is not considered significant. In order to explore the source of the oxygen peaks a deconvolution of the XPS spectrum, for the heat-treated sample and the heat-treated and sonicated, in the binding energy region for C, and O is shown in Figure S3 and Figure S4 respectively. From Figure S3a, the peak at 283.2 eV, the largest peak, is associated with C=C double bonds of the graphene. The remaining C 1s peaks are associated with hydroxyl 284.7 eV. The C 1s peak at 288 .0 eV is associated with the C=O, and the C 1s peak at 289.7 is associated with a COOH/COOR. [46]. From Figure S3b, the O 1s peak at 531.2 eV is associated with COOH, and the O 1s peak at 532.7 eV is associated with the –OH functional group [47]. The deconvolution of the heat-treated and sonicated samples is shown in Figure S4. Similar to the analysis for the heat-treated samples, the peak at 283.2 eV (the largest peak in the spectrum) is associated with C=C double bonds of the graphene. The remaining C1s peak at 284.7 eV is associated with the -OH functional group. Likewise, the C 1s peak at 289.2 eV is associated with COOH/COOR functional groups (see Figures S4a and S4b). In Figure S4b, the O 1s peak at 531.0 eV is associated with COOH, and the O 1s peak at the 532.4 eV is associated with the –OH functional group. From the analysis above one concludes that there is very little -OH or -COOH functionalization on the xGnP; furthermore, the sonication process has very little effect on the chemistry, nor does it reduce the particle size.

In order to investigate particle dispersion at various concentrations of xGnP, tensile samples were placed in liquid nitrogen for about 5 minutes and then snapped in half. SEM micrographs of the fracture surface were then used to study the nanoparticle dispersion in the polymer matrix (see Figure 3 and also Figures S5, S6 and S7). In the lower magnification micrographs, the xGnP shows brighter, due to electron interaction with the jagged edges of the xGnP, than the polyurea matrix, example of the xGnP, are highlighted by the arrows. No agglomeration or continuous network of xGnP was found in any of the formulations. Figure S5f, and S7f are higher magnification micrographs (13kX, and 10 kX respectively) of the xGnP. The edges of the individual nano-plates can be seen, suggesting the GnP remained exfoliated through the sample preparation process.

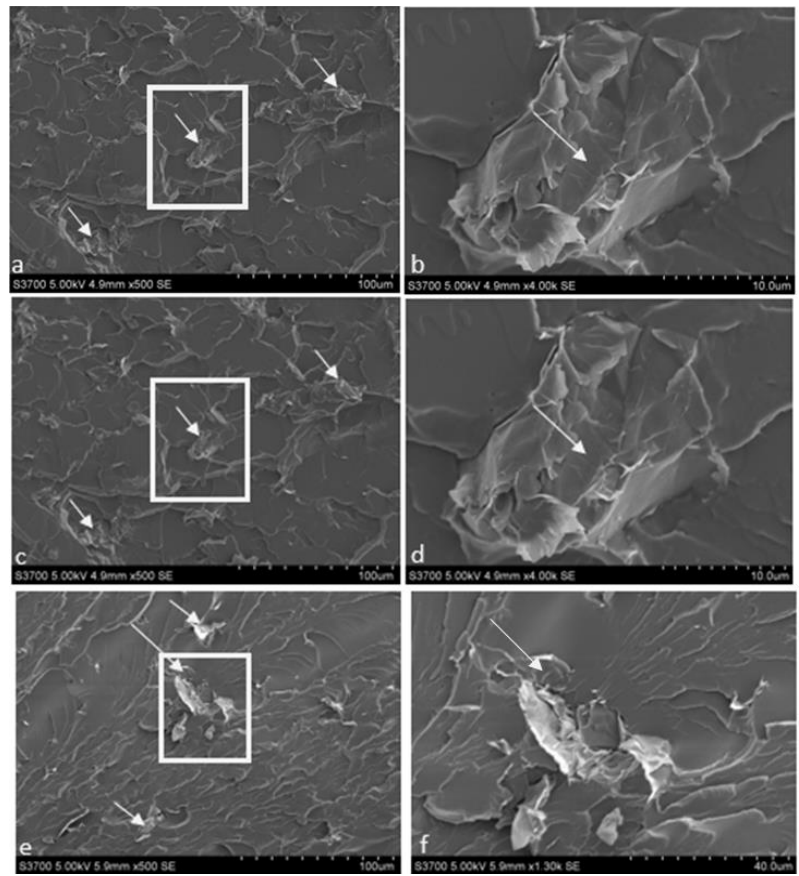

Figure 3 – SEM photomicrographs of the fracture surface for PUa-xGnP nanocomposites with 0.5 wt % xGnP loading: a) 20% HSWF at 500x, b) Photomicrograph of the white box in a. c) 30% HSWF at 500x, b) Photomicrograph of the white box in c. d) 40% HSWF at 650x. Photomicrograph of the white box in e. In all photomicrographs the arrows point to the xGnP.

Figure 4 shows the storage modulus curves for the DMA temperature sweeps (frequency 1 Hz) for the (a) IPDI-2k-20HS, (b) IPDI-2k-30HS, and (c) IPDI-2k-40HS formulation at various xGnP loadings. Note that the complete $E'$, $E''$ and tan($\delta$) curves for all formulations are shown in the supplemental section, Figures S8, S9, and S10. For all formulations, the addition of xGnP did not have an appreciable effect on the $T_g$ (as measured by $E'$, $E''$ or tan($\delta$) curves) of the PUa formulations, nor did it have a significant effect on the glassy modulus. For these formulations, the $T_g$ and glassy modulus are determined primarily by the soft phase. [19] This would tend to indicate the xGnP has little effect on the soft phase microstructure, *i.e.*, no crystallization or increase in the hydrogen bonding in the soft phase. For the IPDI-2k-20HS and IPDI-2k-30HS PUa the addition of xGnP did increase the plateau modulus and the temperature range of the plateau modulus. For the IPDI-2k-40HS formulations additions of xGnP had no effect on the temperature sweeps.

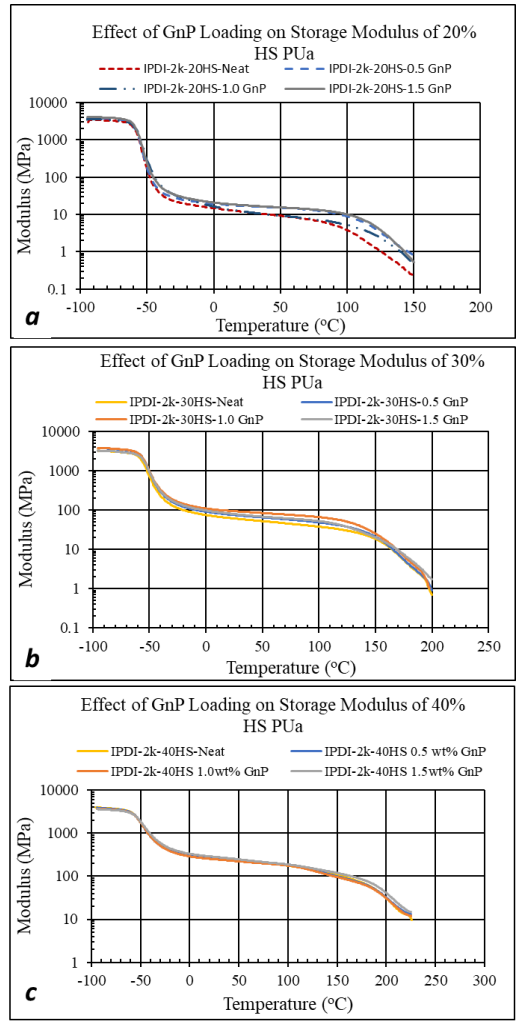

Figure 4. DMA temperature sweeps showing the effect of both increase in % HS and increase in xGnP loading.

### 3.2 Modeling Results

As previously discussed, the linear viscoelastic behavior of polyurea is described by a model comprising two parallel Fractional Maxwell Gel (FMG) branches representing the soft and hard phases. Even though in the nanocomposites, there is a new phase (xGnPs), we continue to use the two-FMG model, and expect that the impact of the nanofillers would be only in modifying the parameters of one or both of the FMGs, at least at sufficiently low loadings (<1.5 wt % in our case). This modeling approach enables us to effectively capture the broad spectrum of relaxation times seen in these materials. The parameterization process was previously detailed, and we now present the results.

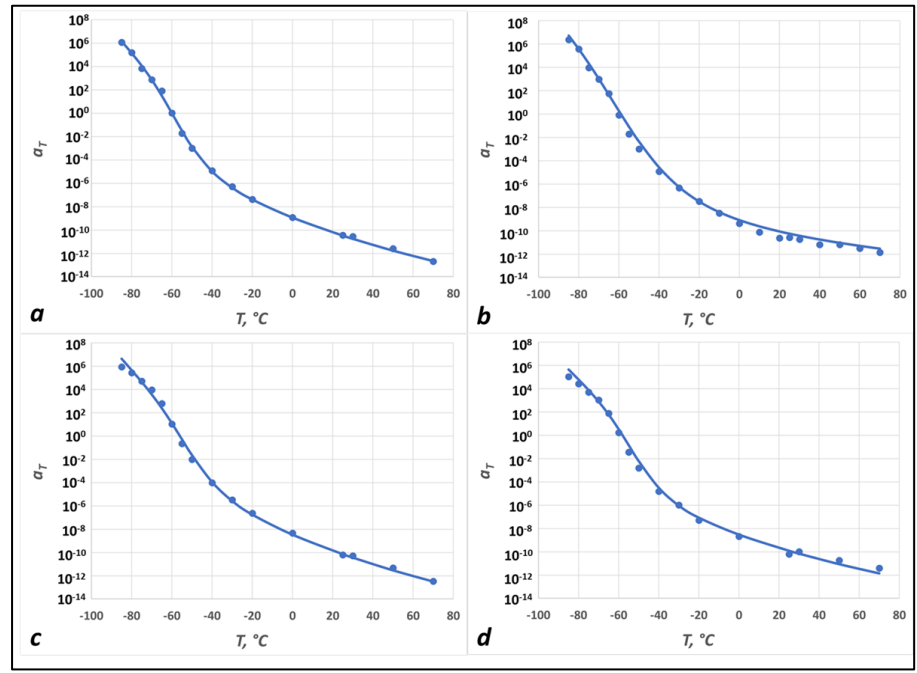

Figure 5. Experimental (symbols) and TS2 fit (lines) shift factors for 20HS polyureas with: (a) No added nanofillers; (b) 0.5 wt % xGnP; (c) 1.0 wt % xGnP; (d) 1.5 wt % xGnP.

To begin with, in Figure 5, we plot the shift factor as a function of temperature for the IPDI-2k-20HS nanocomposites with (a) 0%, (b) 0.5%, (c) 1% and (d) 1.5% xGnP. The symbols are the results of the TTS shift of the data (as outlined above), and the lines are the TS2 (Equation (4)) fits. Obviously, the addition of xGnPs does not have a qualitative impact on the TTS or the temperature dependence of the shift factor, although the model parameters (such as activation energies) do change slightly. Similar data and model fits for IPDI-2k-30HS and IPDI-2k-40HS nanocomposites are presented in Figures S10 – S11, and the TS2 model parameters are summarized in Table S1.

In Figure 6, the storage and loss master curves are plotted for all nanocomposite systems: (a) IPDI-2k-20HS matrix, (b) IPDI-2k-30HS matrix; (c) IPDI-2k-40HS matrix. Within each "family", all curves lie very close to each other, with a possible exception of the IPDI-2k-20HS, 1% xGnP (blue symbols in Figure 6a). We will return to this system later to discuss the origins of its uniqueness.

Next, let us consider the results of the two-FMG fitting to the master curves.

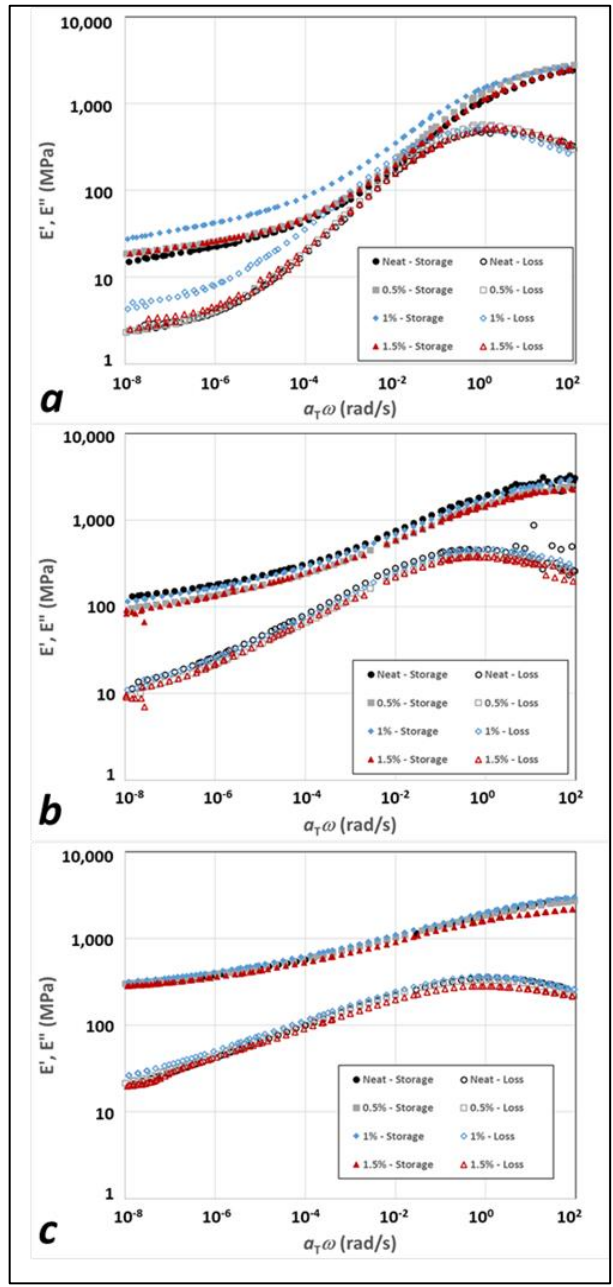

Figure 6. Master curves for the tensile storage (filled symbols) and loss (open symbols) moduli. (a) 20HS matrix with 0, 0.5, 1.0, and 1.5 wt % xGnP. (b) Same as (a) for the 30HS matrix. (c) Same as (a) for the 40HS matrix.

Table 2 provides the mean values and corresponding standard deviations for all six parameters in our two-FMG model. The optimization runs show excellent convergence and reproducibility as manifested in the low standard deviation values for all the systems considered.

Figure 7 presents the two-FMG model fits to the experimental shifted data for IPDI-2k-40HS nanocomposites; the results for IPDI-2k-20HS and IPDI-2k-30HS are depicted in Figures S12 and S13, respectively. All fitted curves are generated using the expected values for the model parameters, since the standard deviation of each model parameter was negligible. For all the formulations, the relative error between model and data was less than 3.1%, with data spanning a broad range of frequencies (between $10^{-4}$ and $10^2$ rad/s).

However, for the 20 wt% hard segment sample at all nano-particle percentages, a minor deviation between the model and experimental data is observed above the glass transition point in the loss modulus, a phenomenon which was also noted in our prior work for the neat 20% HWSF case. It should be noted that the experimental data points exhibiting a high level of dispersion are excluded from the optimization and fitting process.

Figures 8a-b depict the influence of the nanofiller content on the mean characteristic modulus of both branches. In general, the effect is very small, except for the significant increase in $E_{0.2}$ for the 1% xGnP in the IPDI-2k-20HS nanocomposite relative to the neat polymer. In that system, two factors contribute to the effect. First, the stiffness ratio between the filler and the matrix is the largest for the lower-HS polymers and becomes smaller as HSWF is increased. Second, the impact of the fillers usually has a maximum as a function of filler loading. At low loadings, the effect is, obviously, very weak; at the high loadings, on the other hand, the platelets aggregate, the aspect ratio decreases, and the overall effect decreases as well. Thus, 1% xGnP in the IPDI-2k-20HS represents the system corresponding to the maximum reinforcement in terms of both HSWF and %xGnP.

In Figures 8c-d, the variations in relaxation times for both branches with respect to the filler weight fraction are shown. These variations are also fairly small and do not show a clear dependence on the nanoparticle loading. Finally, Figures 8e-f show the power-law exponents, $\alpha$, for both branches. Again, the dependence of $\alpha$ on the xGnP loading is fairly weak. The soft-phase exponent, $\alpha_1$, shows a strong dependence on HSWF, decreasing as HSWF is increased. This is consistent with the material becoming "more elastic" and the average loss tangent decreasing. The hard-phase exponent, $\alpha_2$, is quite small for all twelve neat and nanocomposite systems, indicating that it is almost always nearly perfectly elastic.

Table 2. FMG 1 and FMG 2 parameters. Rows represent hard segment weight fractions of 20%, 30%, and 40%, while columns correspond to xGnP weight fractions of 0.0%, 0.5%, 1%, and 1.5%.

| Model Parameters | | 0.0% GnP | | 0.5% GnP | | 1% GnP | | 1.5% GnP | |
|---|---|---|---|---|---|---|---|---|---|
| | | FMG 1 | FMG 2 | FMG 1 | FMG 2 | FMG 1 | FMG 2 | FMG 1 | FMG 2 |
| **IPDI-2k 20HS** | $E_{0,i}$ (MPa) | 2815 $\pm\,3.3 \times 10^{-6}$ | 64 $\pm\,3.8 \times 10^{-7}$ | 3097 $\pm\,3.4 \times 10^{-6}$ | 51 $\pm\,3.9 \times 10^{-7}$ | 2871 $\pm\,2.6 \times 10^{-6}$ | 118 $\pm\,6.2 \times 10^{-7}$ | 2837 $\pm\,2.6 \times 10^{-6}$ | 71 $\pm\,4.1 \times 10^{-7}$ |
| | $\tau_{c,i}$ (s) | 0.18 $\pm\,8.2 \times 10^{-10}$ | 1.19 $\pm\,5.9 \times 10^{-10}$ | 0.22 $\pm\,6.8 \times 10^{-10}$ | 1.69 $\pm\,7.1 \times 10^{-10}$ | 0.40 $\pm\,1.4 \times 10^{-9}$ | 1.96 $\pm\,8.1 \times 10^{-9}$ | 0.18 $\pm\,6.7 \times 10^{-10}$ | 1.16 $\pm\,4.5 \times 10^{-9}$ |
| | $\alpha_i$ | 0.42 $\pm\,4.3 \times 10^{-10}$ | 0.080 $\pm\,3.8 \times 10^{-10}$ | 0.413 $\pm\,4.4 \times 10^{-10}$ | 0.056 $\pm\,5.8 \times 10^{-10}$ | 0.382 $\pm\,3.5 \times 10^{-10}$ | 0.085 $\pm\,3.4 \times 10^{-10}$ | 0.412 $\pm\,4.0 \times 10^{-10}$ | 0.077 $\pm\,3.9 \times 10^{-10}$ |
| **IPDI-2k 30HS** | $E_{0,i}$ (MPa) | 2757 $\pm\,1.8 \times 10^{-5}$ | 342 $\pm\,1.7 \times 10^{-6}$ | 2405 $\pm\,3.4 \times 10^{-6}$ | 278 $\pm\,1.5 \times 10^{-6}$ | 2793 $\pm\,2.5 \times 10^{-6}$ | 341 $\pm\,1.2 \times 10^{-6}$ | 2203 $\pm\,1.9 \times 10^{-6}$ | 301 $\pm\,8.9 \times 10^{-6}$ |
| | $\tau_{c,i}$ (s) | 1.14 $\pm\,4.1 \times 10^{-8}$ | 3.23 $\pm\,1.0 \times 10^{-7}$ | 0.68 $\pm\,5.8 \times 10^{-9}$ | 2.00 $\pm\,1.2 \times 10^{-8}$ | 0.66 $\pm\,2.7 \times 10^{-9}$ | 1.90 $\pm\,6.3 \times 10^{-9}$ | 0.89 $\pm\,2.7 \times 10^{-8}$ | 2.41 $\pm\,3.0 \times 10^{-8}$ |
| | $\alpha_i$ | 0.314 $\pm\,1.1 \times 10^{-9}$ | 0.054 $\pm\,3.6 \times 10^{-10}$ | 0.307 $\pm\,5.7 \times 10^{-10}$ | 0.061 $\pm\,3.4 \times 10^{-10}$ | 0.307 $\pm\,3.4 \times 10^{-10}$ | 0.061 $\pm\,2.2 \times 10^{-10}$ | 0.312 $\pm\,2.5 \times 10^{-9}$ | 0.071 $\pm\,2.0 \times 10^{-9}$ |
| **IPDI-2k 40HS** | $E_{0,i}$ (MPa) | 2635 $\pm\,2.6 \times 10^{-6}$ | 604 $\pm\,2.0 \times 10^{-6}$ | 2476 $\pm\,2.0 \times 10^{-6}$ | 622 $\pm\,1.7 \times 10^{-6}$ | 2843 $\pm\,2.8 \times 10^{-6}$ | 687 $\pm\,2.8 \times 10^{-6}$ | 2567 $\pm\,1.3 \times 10^{-5}$ | 378 $\pm\,1.1 \times 10^{-5}$ |
| | $\tau_{c,i}$ (s) | 0.69 $\pm\,2.8 \times 10^{-9}$ | 1.44 $\pm\,5.2 \times 10^{-9}$ | 1.03 $\pm\,3.4 \times 10^{-9}$ | 2.06 $\pm\,8.8 \times 10^{-9}$ | 0.46 $\pm\,2.1 \times 10^{-9}$ | 0.93 $\pm\,4.9 \times 10^{-9}$ | 0.17 $\pm\,2.7 \times 10^{-9}$ | 0.45 $\pm\,3.7 \times 10^{-9}$ |
| | $\alpha_i$ | 0.242 $\pm\,4.1 \times 10^{-10}$ | 0.035 $\pm\,1.8 \times 10^{-10}$ | 0.239 $\pm\,2.8 \times 10^{-10}$ | 0.038 $\pm\,1.6 \times 10^{-10}$ | 0.222 $\pm\,3.4 \times 10^{-10}$ | 0.043 $\pm\,1.8 \times 10^{-10}$ | 0.187 $\pm\,1.2 \times 10^{-9}$ | 0.015 $\pm\,1.7 \times 10^{-9}$ |

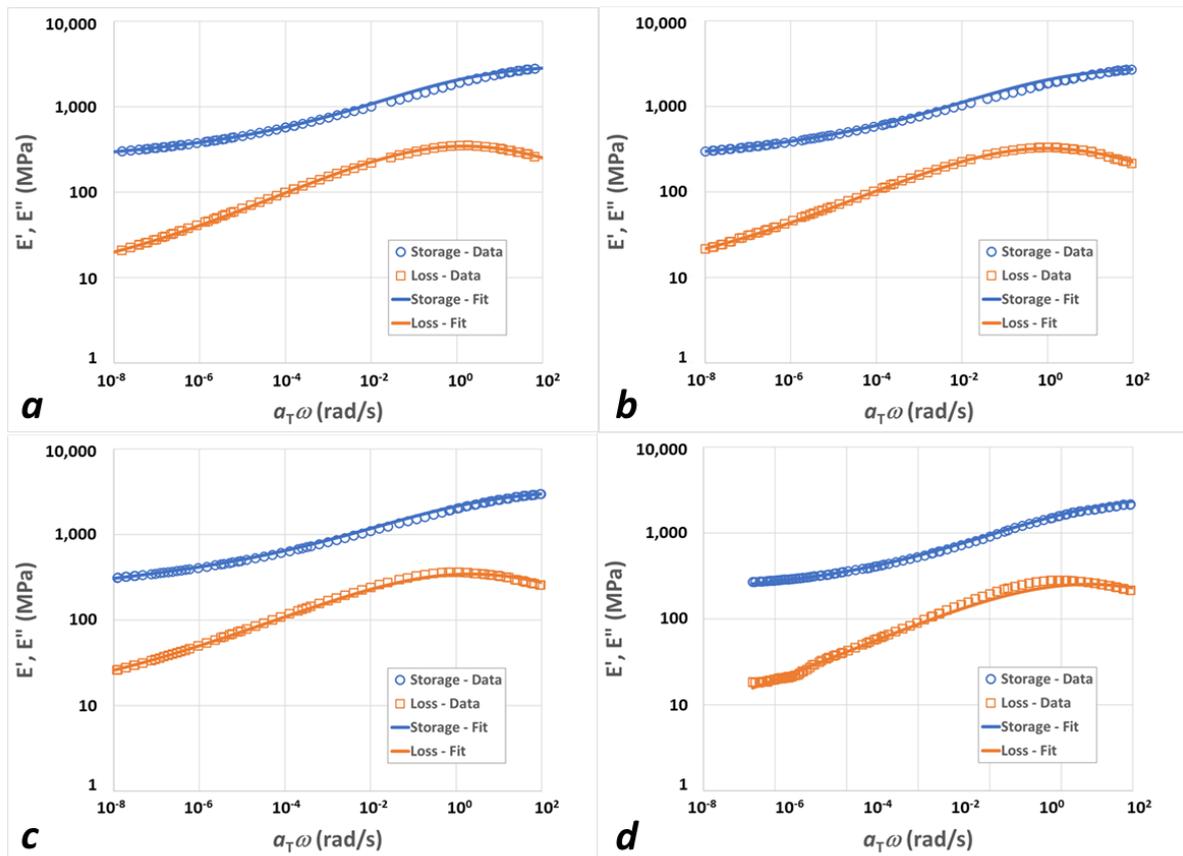

Figure 7. Experimental (symbols) and FMG-FMG fit (lines) master curves for 40HS polyureas with: (a) No added nanofillers; (b) 0.5 wt % xGnP; (c) 1.0 wt % xGnP; (d) 1.5 wt % xGnP. Blue open circles represent storage modulus data, blue lines are the storage modulus model fits; orange open squares correspond to the loss modulus data, and orange lines are the loss modulus model fits.

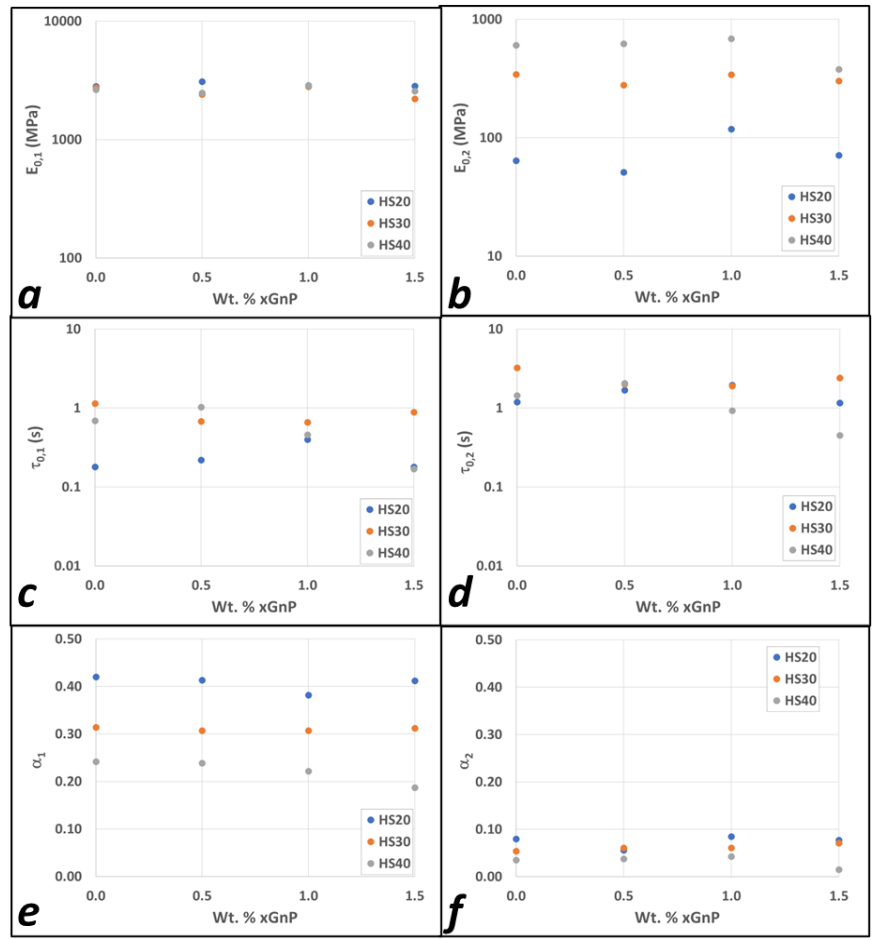

Figure 8. Effect of the nanofiller loading on the (a) characteristic modulus of the first branch ($E_{0,1}$) (b) characteristic modulus of the second branch ($E_{0,2}$), (c) characteristic time of the first branch ($\tau_{c,1}$), (d) characteristic time of the second branch ($\tau_{c,2}$), (e) power-law exponent of the first branch ($\alpha_1$), and (f) power-law exponent of the first branch ($\alpha_2$).

## 4. Discussion

In this study, we investigated the structure and linear viscoelasticity of polyurea elastomers to be used in adhesive applications. The two main variables of interest were the polyurea hard segment (HS) weight fraction and the exfoliated graphene nanoplatelet (xGnP) loading. The hypothesis tested was that the polyurea hard segment and the nanofillers interact strongly with each other and provide additional reinforcement by forming a "combined hard phase".

Using Scanning Electron Microscopy (SEM), we verified that the heat-treatment and the sonication in toluene resulted in no morphological changes to the xGnP. The average particle diameter did not change and the xGnP's remained exfoliated and well dispersed in the PUa matrix. Recall that the nanoparticles were placed on the Isocyanate side (A-side) of the PUa reaction sequence, therefore, albeit small, there is a potential for the isocyanate to react with any hydroxyl or carboxylic acid functional groups located on the edges of the nano particles. However, XPS showed very few, if any, available reaction sites, whether they be hydroxyl or carboxylic acid that could potential react with the isocyanate. Thus, we stipulate that the dispersed xGNPs have only weak physical interactions with the PUa matrix.

Given the complex structure of any polyurea nanocomposite (soft-phase matrix, hard-phase islands, percolated hard-phase domains, exfoliated nanofillers, aggregated nanofillers, etc.), the data from direct characterization such as electron microscopy is often inconclusive. Thus, here we also concentrated on understanding the materials using linear viscoelasticity and inferring the information about the matrix-filler interaction from the DMA results.

Similar to the previous study, [19] we observed that the DMA frequency sweeps in these systems are amenable to time-temperature superposition (TTS), with the TTS shift factors well-described by the TS2 [22] function. This was, in itself, a non-trivial result, since polyurea materials are multi-phase; understanding the reason why TTS works still requires additional analysis. We also found that the storage and loss master curves exhibit broad transition regions and thus cannot be described with a single Maxwell model. Therefore, we used the fractional Maxwell model (FMM) approach [23–27] to quantify the viscoelastic response and fit the master curve. In particular, the material was well-described by use of two fractional Maxwell gel (FMG) elements, one representing the soft phase, and another one the percolated hard phase. We demonstrated that the plateau modulus of the percolated hard phase (FMG2) increased strongly with the hard segment weight fraction (HSWF), consistent with earlier studies.[10] Here, we used the same approach to determine the combined impact of HSWF and xGnP loading.

Based on the FMM analysis, we observed that the effect of the xGnP is significantly less pronounced than the effect of the HSWF change. The reinforcement factor (RF) is physically meaningful (significantly greater than 1) only for one nanocomposite system – HS20 with 1% xGnP. This result is consistent with expectations, as discussed above. Further increases in the xGnP loading likely result in at least some aggregation, thus blunting the effectiveness of the new fillers.[48] For polyureas with higher HSWF, the percolated hard phase modulus is already quite high, and the contribution of the nanofillers becomes even less significant, regardless of their concentration. Thus, the addition of the nanofillers does not seem to offer significant increase in the linear elastic properties of the polyureas studied here.

Of course, linear elasticity is not the only important property for adhesives – other properties of interest include tensile strength, ultimate elongation, fracture toughness, etc. The influence of nanofillers on those properties will be the subject of future work.

## 5. Conclusions

We investigated the structure and linear viscoelasticity of polyurea (PUa) elastomers and their nanocomposites with expanded graphene nanoplatelets (xGNP) as function of the hard segment weight fraction (HSWF) of the polyurea and the xGNP weight fraction in the overall nanocomposites. Experimentally, we found that the room-temperature modulus of the PUa-xGNP nanocomposites depends strongly on HSWF (about 10 MPa for the 20% HSWF to about 100 MPa for the 30% HSWF to about 250 MPa for the 40% HSWF polymer), but weakly on the xGNP weight fraction (for the weight fraction variations between 0 and 1.5 wt %, the modulus variations were generally within the experimental error, except for the 20%HSWF, 1% xGNP nanocomposite exhibiting nearly two-fold stiffening compared to the neat material).

Significantly, we have demonstrated that despite their structural complexity, PUa-xGNP nanocomposites exhibit time-temperature superposition (TTS). For the first time, we demonstrated that the TTS master curves can be described by the fractional calculus (FC) based models with a small number of physically meaningful parameters (as opposed to the standard Prony series modeling usually requiring twenty or more). The new model can be adapted to describe other polymers and nanocomposites for both linear and non-linear mechanical tests.

Supplementary Materials for:

# Polyurea-Graphene Nanocomposites – the Influence of Hard-Segment Content and Nanoparticle Loading on Mechanical Properties

**Demetrios A. Tzelepis [1], Arman Khoshnevis [2], Mohsen Zayernouri [3] and Valeriy V. Ginzburg [4*]**

[1]  Department of Chemical Engineering and Materials Science, Michigan State University, East Lansing, Michigan, USA; email tzelepi1@msu.edu

[2]  Department of Mechanical Engineering, Michigan State University, East Lansing, Michigan, USA; email khoshne1@msu.edu

[3]  Department of Mechanical Engineering, Michigan State University, East Lansing, Michigan, USA; email zayern@msu.edu

[4]  Department of Chemical Engineering and Materials Science, Michigan State University, East Lansing, Michigan, USA; email ginzbur7@msu.edu

[*]  Correspondence: ginzbur7@msu.edu

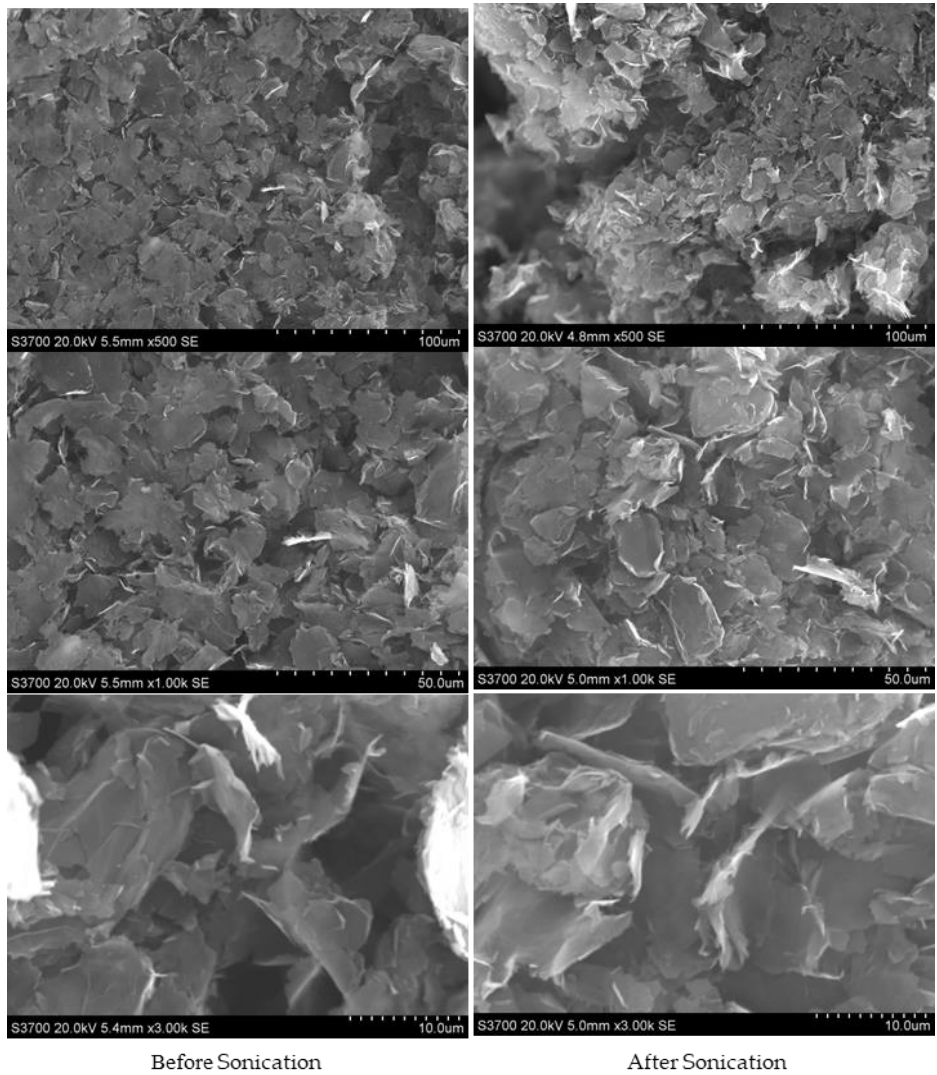

Before Sonication          After Sonication

Figure S1 - SEM images at various magnifications of the xGnP before (left column) and after sonication (right column). The images indicate that there is no change in the morphology of the xGnP with the sonication parameters used.

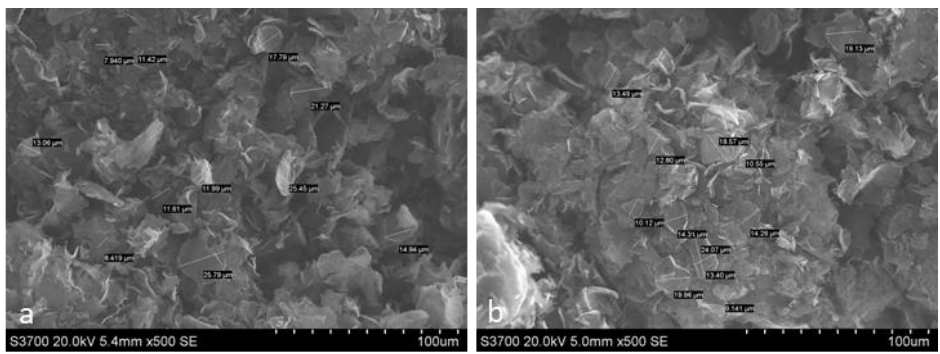

Figure S2 - SEM micrographs showing the measurement of the estimated diameter of the xGnP before sonication (a) and after sonication (b).

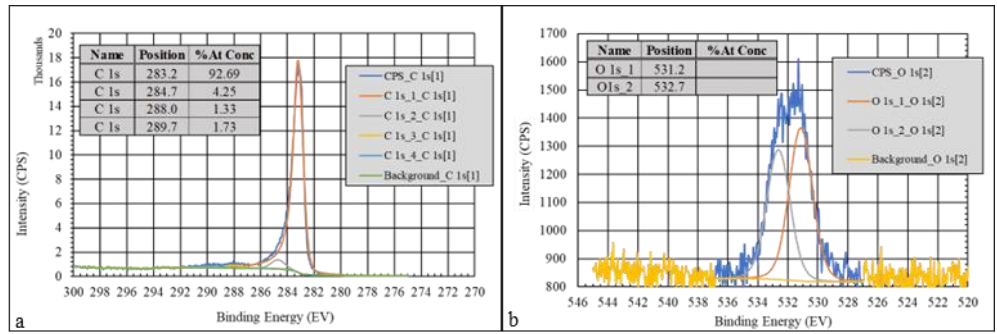

Figure S3 – Deconvolution of the XPS spectrum, for the heat-treated xGnP, in the binding energy region for C (a), and O (b).

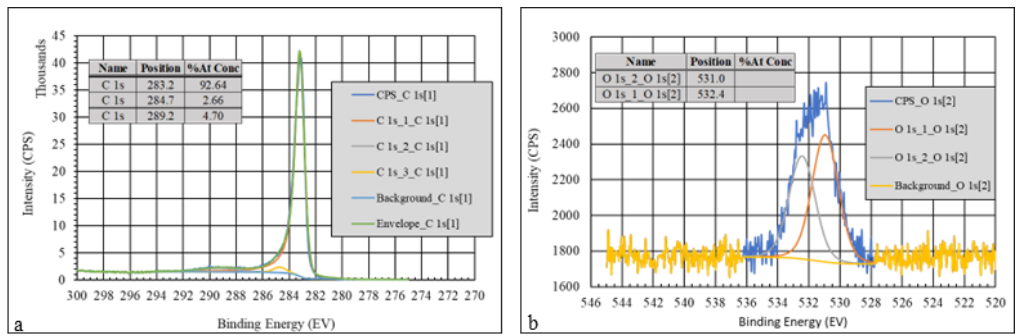

Figure S4 - Deconvolution of the XPS spectrum, for the heat-treated and sonicated xGnP and sonicated, in the binding energy region for C (a), and O (b).

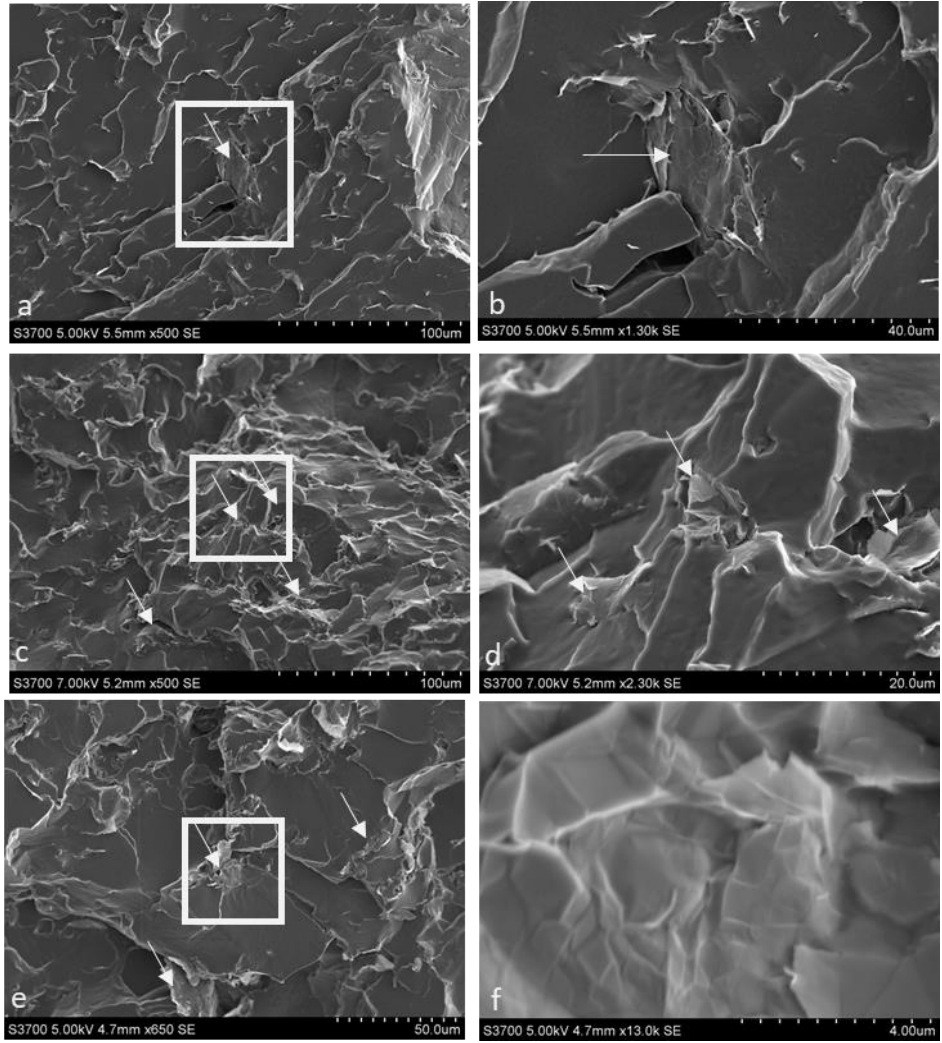

Figure S5 - SEM photomicrographs of the fracture surface for all IPDI-2k-20HS xGnP loadings: a) 0.5 wt% xGnP loading at 500x, b) Photomicrograph of the white box in a. c) 1.0 wt% xGnP loading at 500x, d) Photomicrograph of the white box in c. e) 1.5 wt% xGnP loading; f) Photomicrograph of the white box in e.

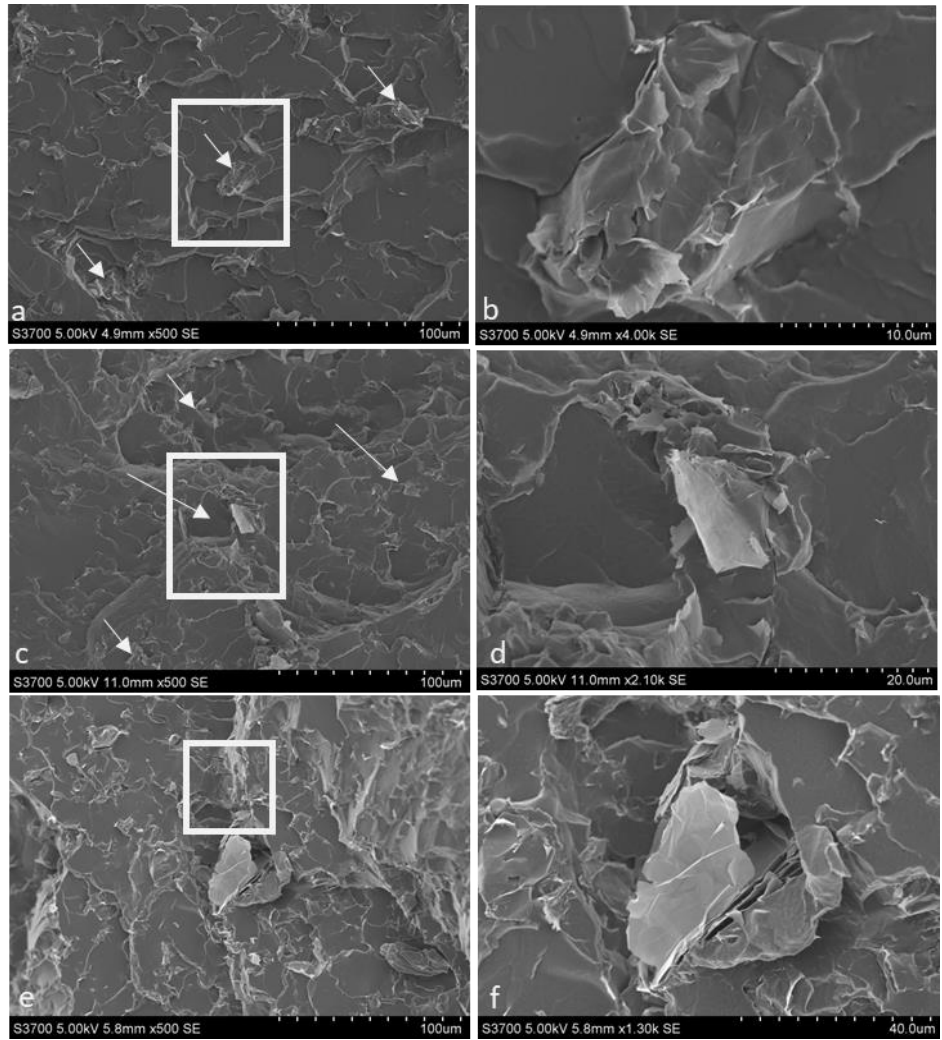

Figure S6 - SEM photomicrographs of the fracture surface for all IPDI-2k-30HS xGnP loadings: a)    0.5 wt% xGnP loading at 500x, b) Photomicrograph of the white box in a. c) 1.0 wt% xGnP loading at 500x, d) Photomicrograph of the white box in c.    e) 1.5 wt% xGnP loading at 650x. f) Photomicrograph of the white box in e.    In all photomicrographs the arrows point to the xGnP.

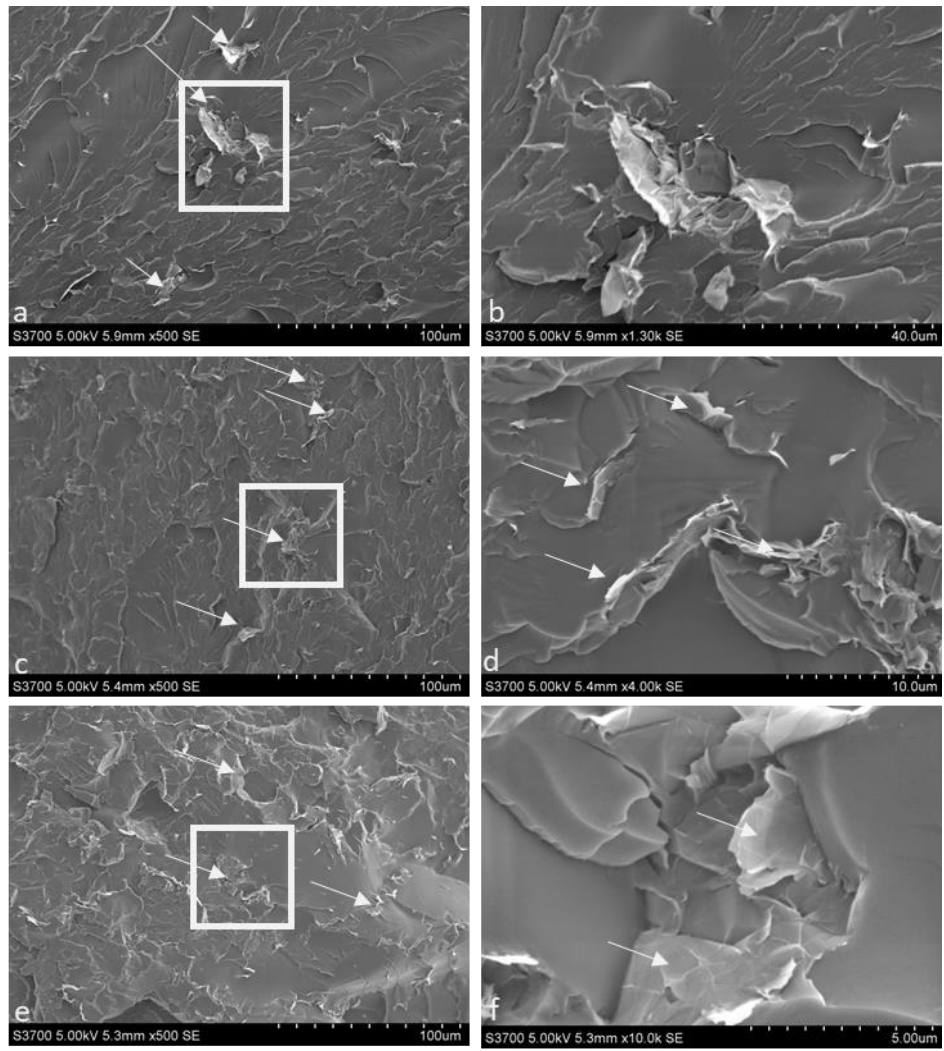

Figure S7 - SEM photomicrographs of the fracture surface for all IPDI-2k-40HS xGnP loadings: a) 0.5 wt% xGnP loading at 500x, b) Photomicrograph of the white box in a. c) 1.0 wt% xGnP loading at 500x, d) Photomicrograph of the white box in c. e) 1.5 wt% xGnP loading at 650x. f) Photomicrograph of the white box in e. In all photomicrographs the arrows point to the xGnP.

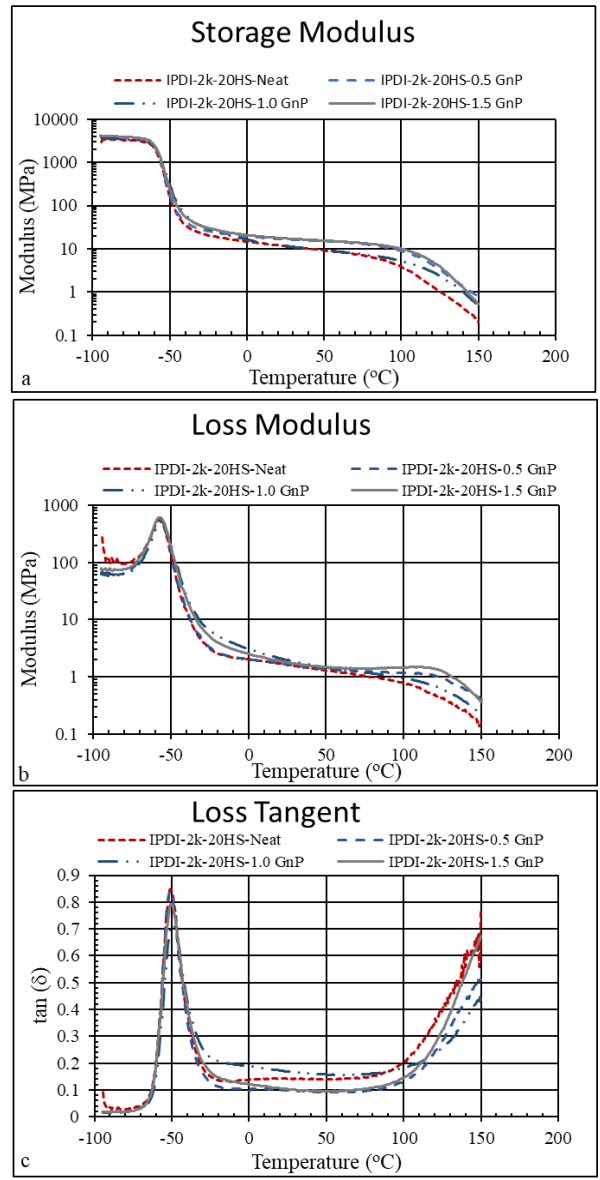

Figure S8 – DMA temperature sweep showing the storage, loss modulus and the tan (δ) for the IPDI-2k-20HS.

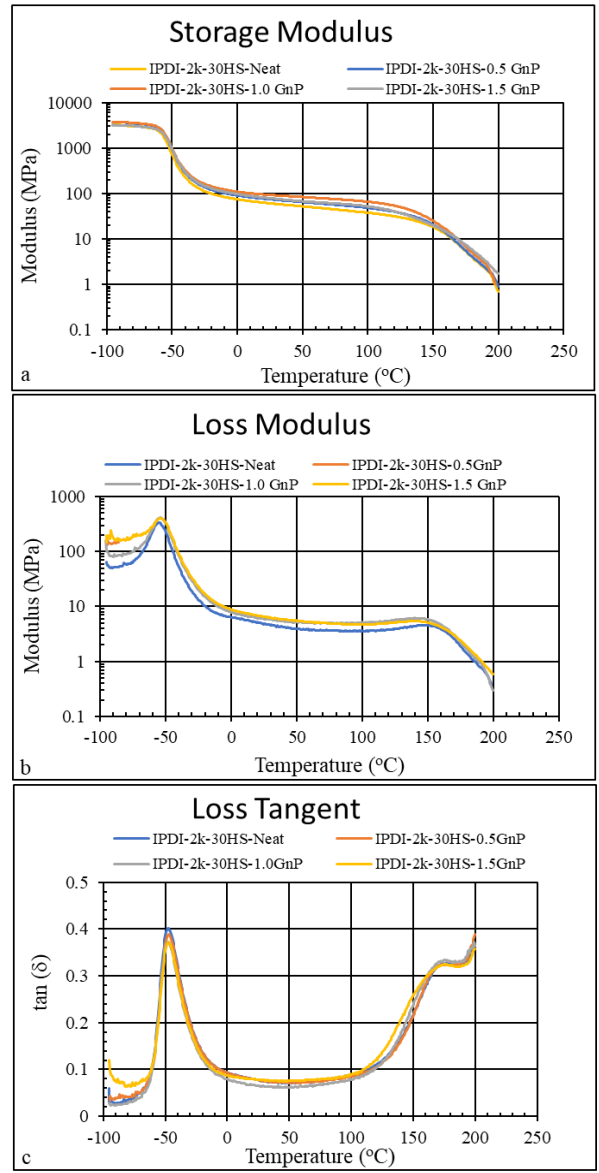

Figure S9 – DMA temperature sweep showing the storage, loss modulus and the tan (δ) for the IPDI-2k-30HS.

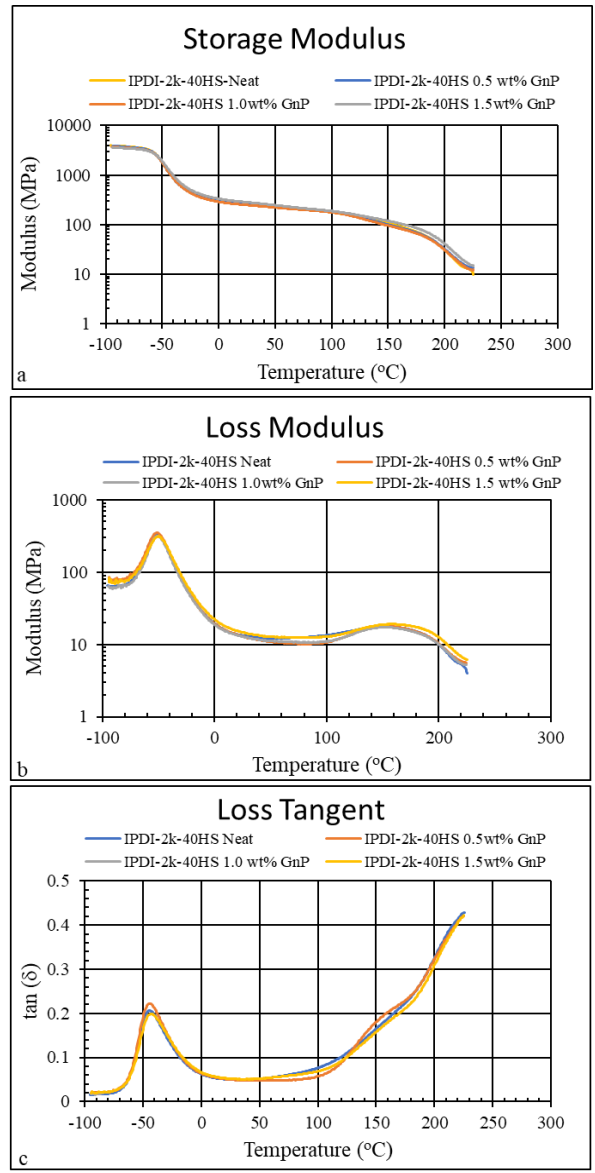

Figure S10 – DMA temperature sweep showing the storage, loss modulus and the tan (δ) for the IPDI-2k-40HS.

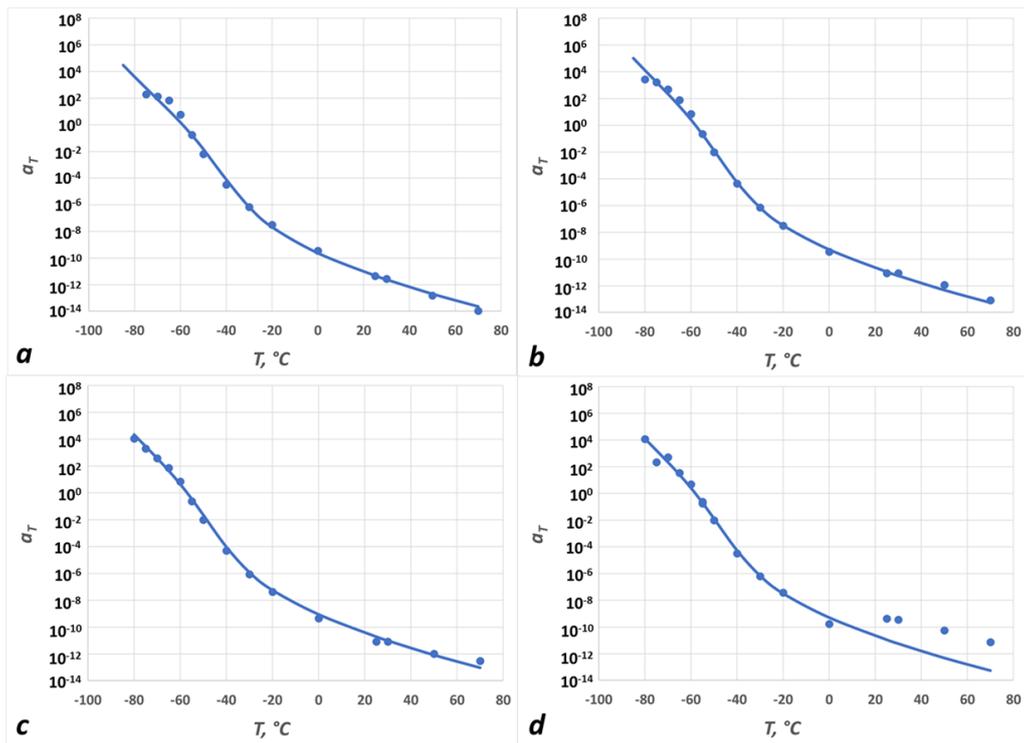

Figure S11: Experimental (symbols) and TS2 fit (lines) shift factors for IPDI-2k-30HS polyureas with: (a) No added nanofillers; (b) 0.5 wt % xGnP; (c) 1.0 wt % xGnP; (d) 1.5 wt % xGnP.

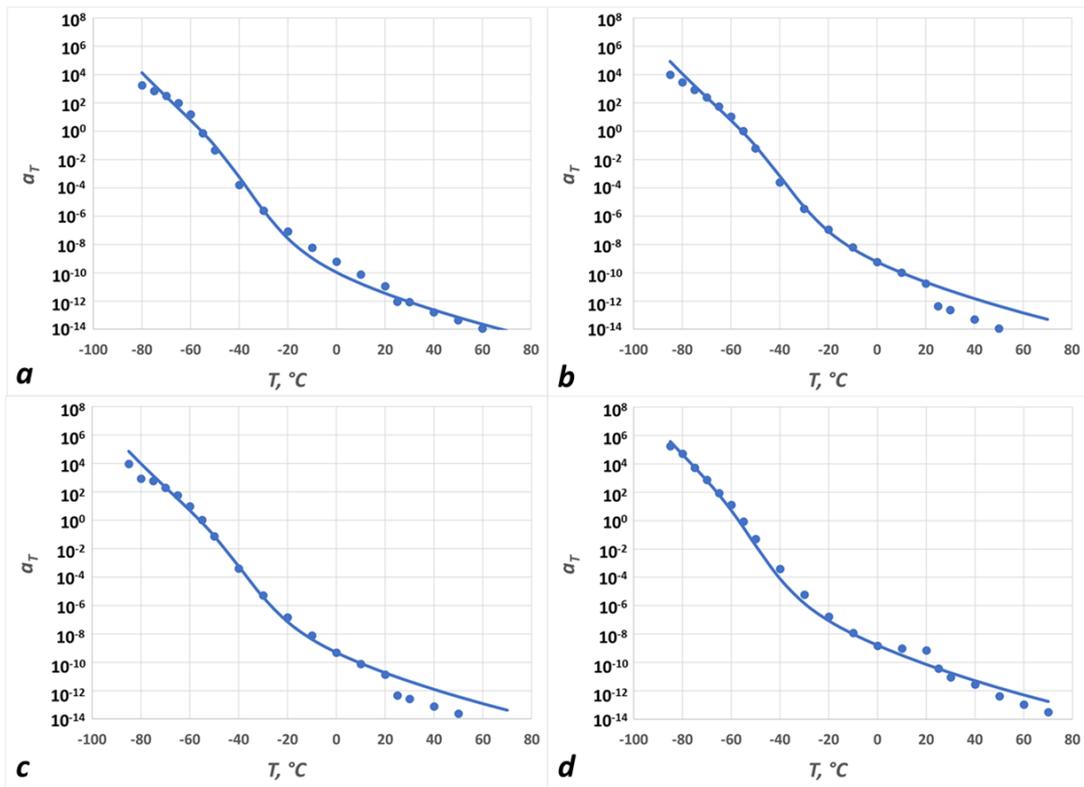

Figure S12: Experimental (symbols) and TS2 fit (lines) shift factors for 40HS polyureas with: (a) No added nanofillers; (b) 0.5 wt % xGnP; (c) 1.0 wt % xGnP; (d) 1.5 wt % xGnP

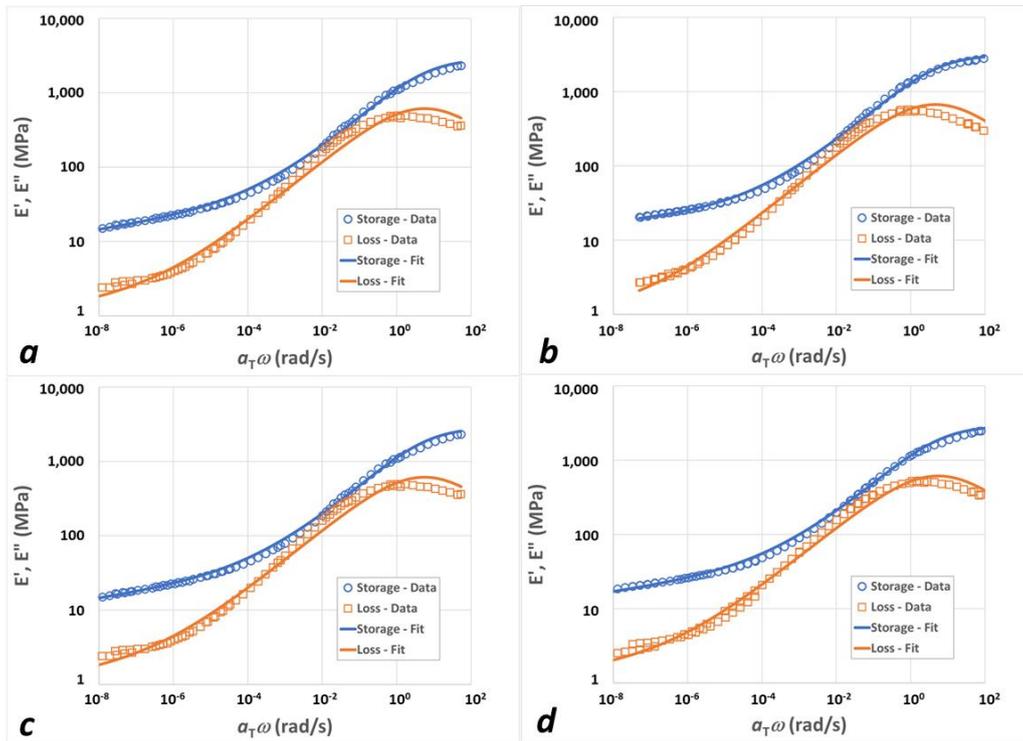

Figure S13: Experimental (symbols) and FMG-FMG fit (lines) master curves for IPDI-2k-20HS polyureas with: (a) No added nanofillers; (b) 0.5 wt % xGnP; (c) 1.0 wt % xGnP; (d) 1.5 wt % xGnP. Blue open circles represent storage modulus data, blue lines are the storage modulus model fits; orange open squares correspond to the loss modulus data, and orange lines are the loss modulus model fits

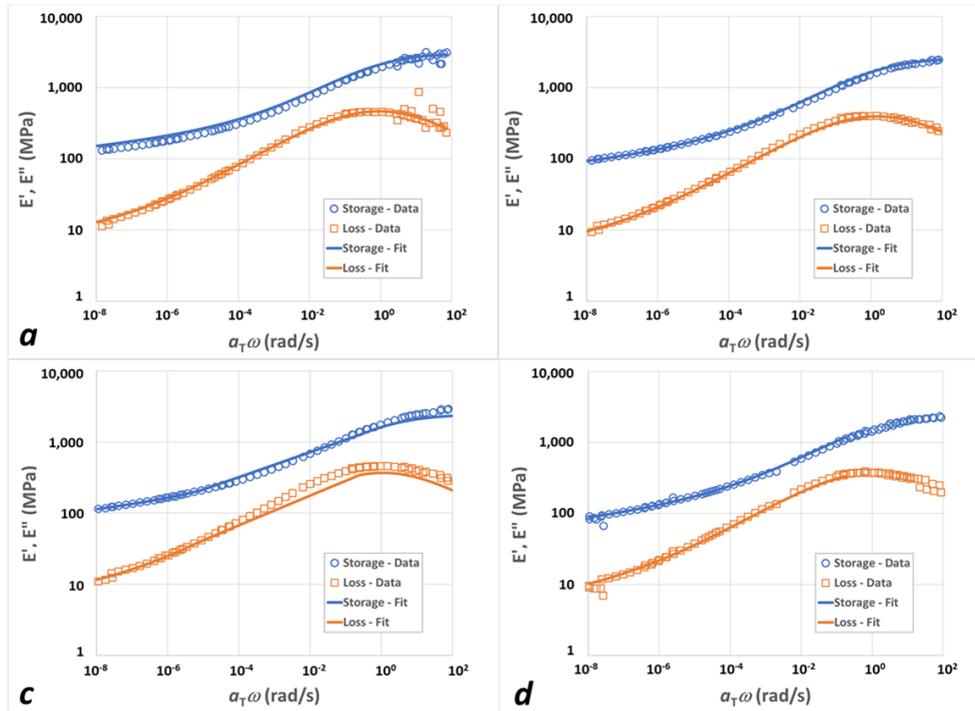

Figure S14: Experimental (symbols) and FMG-FMG fit (lines) master curves for 30HS polyureas with: (a) No added nanofillers; (b) 0.5 wt % xGnP; (c) 1.0 wt % xGnP; (d) 1.5 wt % xGnP. Blue open circles represent storage modulus data, blue lines are the storage modulus model fits; orange open squares correspond to the loss modulus data, and orange lines are the loss modulus model fits

Table S1: TTS reference temperatures and TS2 fit parameters for all systems

| Polymer | %xGnP | $T_0$, K | $E_1$, kJ/mol | $E_2$, kJ/mol | $\Delta S/R$ | $T^*$, K |
|---|---|---|---|---|---|---|
| **IPDI-2k-20HS** | 0.00% | 213 | 94.5 | 120.0 | 25.0 | 217.1 |
| | 0.50% | 214 | 94.5 | 121.1 | 25.0 | 218.5 |
| | 1.00% | 217 | 94.5 | 118.8 | 25.0 | 222.5 |
| | 1.50% | 215 | 94.5 | 116.5 | 25.0 | 222.8 |
| **IPDI-2k-30HS** | 0.00% | 214 | 99.9 | 120.0 | 25.0 | 233.8 |
| | 0.50% | 215 | 99.9 | 120.6 | 25.0 | 228.9 |
| | 1.00% | 216 | 99.9 | 120.6 | 25.0 | 228.9 |
| | 1.50% | 215 | 99.9 | 120.6 | 25.0 | 228.9 |
| **IPDI-2k-40HS** | 0.00% | 217 | 100.0 | 123.6 | 25.0 | 240.5 |
| | 0.50% | 218 | 100.0 | 120.4 | 25.0 | 238.5 |
| | 1.00% | 217 | 100.0 | 120.4 | 25.0 | 238.5 |
| | 1.50% | 216 | 100.0 | 120.8 | 25.0 | 225.6 |